\documentclass[amsmath, twocolumn, showpacs, aps, prb]{revtex4}
\usepackage{bm}
\usepackage{graphics}
\usepackage{graphicx}
\usepackage{color,bigints}

\newcommand{\bnabla}{{\boldsymbol \nabla}}

\newcommand{\brho}{{\boldsymbol \rho}}

\begin{document}

\title{Correlations between Majorana fermions through a superconductor}
\author{A.A. Zyuzin, Diego Rainis, Jelena Klinovaja, and Daniel Loss}

\affiliation{Department of Physics, University of Basel,
Klingelbergstrasse 82, CH-4056 Basel, Switzerland}


\begin{abstract}
We consider a model of ballistic quasi-one dimensional semiconducting wire with intrinsic spin-orbit interaction placed on the surface of a bulk $s$-wave superconductor (SC), in the presence of an external magnetic field. 
This setup has been shown to give rise to a topological superconducting state in the wire, characterized by a pair of Majorana-fermion (MF) bound states formed at the two ends of the wire. 
Here we demonstrate that, besides the well-known direct overlap-induced energy splitting, the two MF bound states may hybridize via elastic tunneling processes through virtual quasiparticles states in the SC, 
giving rise to an additional energy splitting between MF states from the same as well as from different wires.
\end{abstract}
\maketitle

The topological properties of  Majorana fermions (MFs)  as well as the 
search for suitable solid-state experimental scenarios where MFs can arise have been 
attracting wide attention in the last decade~\cite{bib:WilczekReview, bib:AliceaReview, bib:FranzReview}, spurred even more by recent experiments
investigating MF physics \cite{bib:Exp1, bib:Exp2, bib:Exp3, bib:fractionalJ, bib:unconventionalJ, bib:Exp4, bib:Exp1Contr, bib:Exp2Contr}.

There is a number of proposed systems that might host MFs, such as 
two-dimensional spinless chiral superconductors (SCs) with $p+ip$ symmetry including 
the $^3$\textrm{He} A-phase and 
the fractional quantum Hall liquid at filling factor $5/2$ \cite{bib:Volovik1, bib:Volovik-Review, bib:Read, bib:Nayak};
the superfluid $^3$\textrm{He} B-phase \cite{bib:Salomaa, bib:Zhang, bib:Volovik1};
heterostructures of topological insulators and SCs \cite{bib:Fu, bib:Tanaka}; optical lattices \cite{bib:Sato};
quasi one-dimensional systems with spin-orbit interactions (SOI) and proximity induced superconductivity for semiconducting nanowires \cite{bib:Lutchyn, bib:Oreg, bib:AliceaPRB} and for carbon-based materials 
 \cite{bib:MF_CNT_2012, bib:bilayer_MF_2012, bib:nanoribbon_KL}.

Here we consider a model of a wire with Rashba SOI  brought into contact with a bulk $s$-wave SC.
Due to the interplay of proximity-induced %$s$-wave 
superconductivity, SOI, and magnetic field, the wire is expected to enter  a topological superconducting (TSC) phase for strong enough magnetic fields, and to host MF bound states localized at its two ends~\cite{bib:Lutchyn, bib:Oreg, bib:AliceaPRB}.

The energy of an isolated MF is pinned to the Fermi level inside the mini gap, due to its topological nature. 
In realistic finite-size wires the two end MF wave functions overlap, and such coupling leads to the splitting in energy of the otherwise doubly degenerate level. 
In most theoretical approaches, after one has calculated the proximity-induced gap in the wire, one usually forgets about the bulk SC and works with an effective model for the wire.
The smallness of the energy splitting of the MF state, relevant for quantum computing purposes, is then determined by the relation between the wire length $L$ and the MF localization length $\xi_{\rm w}$.
Namely, in order to have an exponentially small splitting, it is necessary to require $L\gg\xi_{\rm w}$.

However, we show here that coupling between MFs can be established also through the SC, on a relevant length scale dictated by the coherence length $\xi_{\rm s}$ {\it in the SC} (modified by inverse power-law corrections in $L$).
In such cases, the energy splitting is exponentially suppressed in the regime $L\gg\xi_{\rm s}$. This SC-mediated effect becomes significant if $\xi_{\rm s}>\xi_{\rm w}$, and, together with other decoherence mechanisms~\cite{bib:Chamon,bib:Trauzettel,bib:Diego3,bib:Diego2}, it could become an important issue. On the other hand, this effect also provides useful signatures that can help to identify MFs experimentally.

Generally speaking,  tunneling between normal %metal or polarized ferromagnetic 
leads via an $s$-wave SC can  occur via elastic single-electron cotunneling processes %\cite{bib:Nazarov}
or via local or crossed  Andreev reflection~\cite{bib:Hekking,bib:Deutscher,bib:Falci}, which is a two-particle tunneling process.
\begin{figure}[t]  \centering
\includegraphics[width=0.9\linewidth] {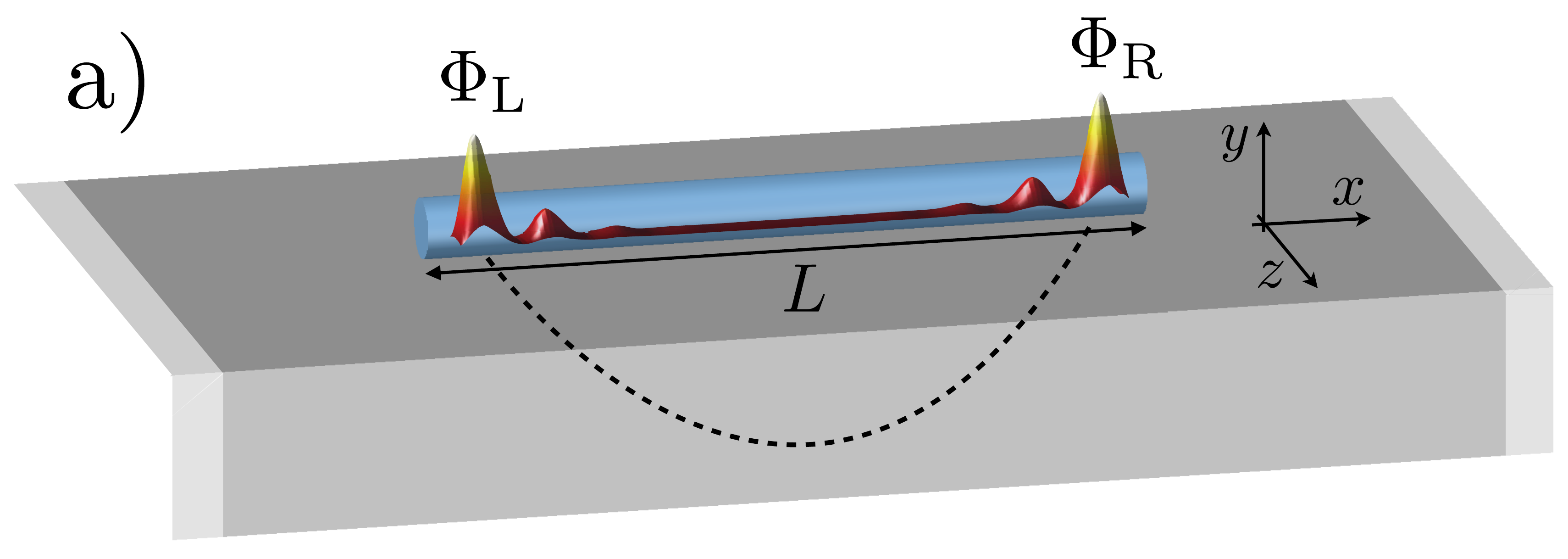}
\includegraphics[width=0.9\linewidth] {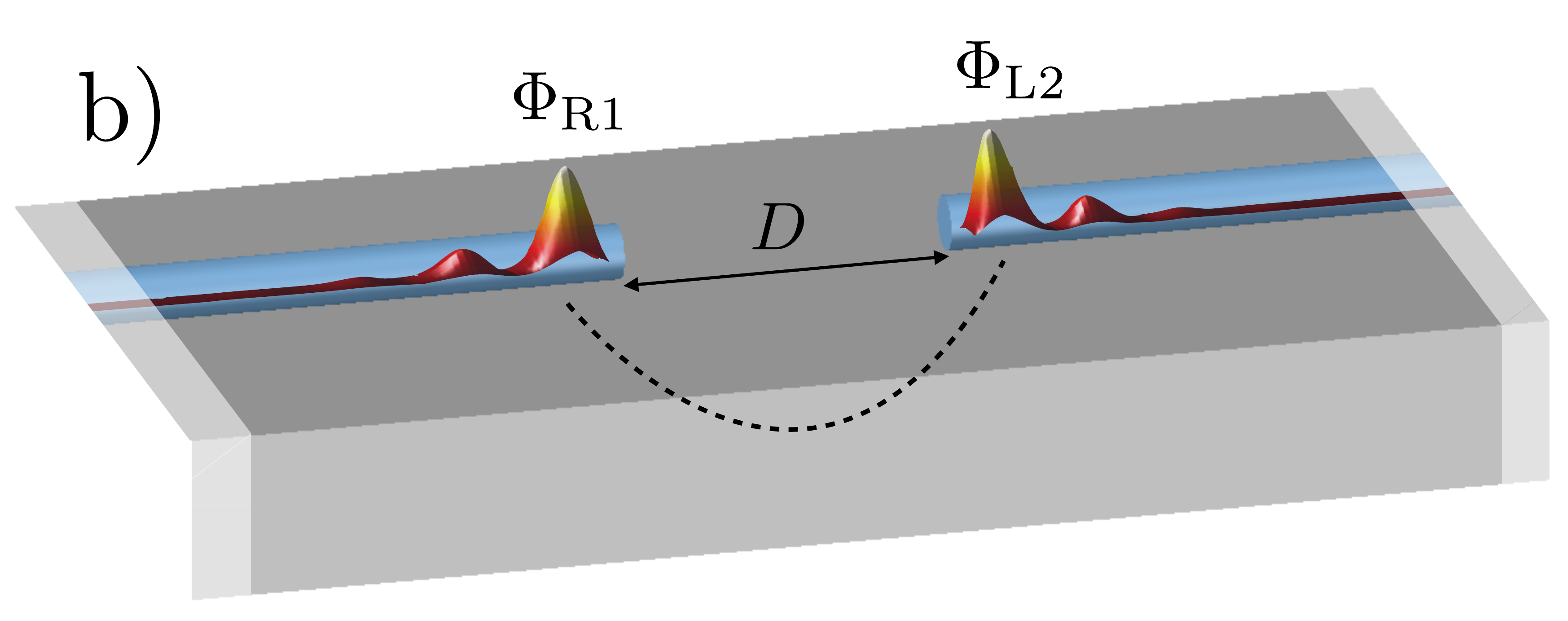}
\caption{ (a) A semiconducting wire (blue) is in tunneling contact with a bulk SC (gray). 
The wire length $L$ is of the order of the coherence length $\xi_{\rm s}$ of the SC.
Two MFs arise in the topological phase and are localized at the wire ends (red). 
The dashed line indicates possible tunneling processes between the two MFs that can couple them and lift their degeneracy.
(b) Setup with two wires, with an inter-wire separation $D\sim \xi_{\rm s}$. The same process can occur between MFs of adjacent wire ends, where now an additional phase difference modulates their coupling.}\label{fig:1}
\end{figure}
Tunneling in systems with MFs are different in that sense.
Two MF states form a {\it single} complex fermionic state 
and coupling via the anomalous propagator of the SC is thus not possible. 
Still, we show here that hybridization between two MFs can be induced by 
coherent tunneling of electrons via virtual quasiparticle states of the SC. 
 
{\it Model.---}%
The setups considered here are schematically shown in Fig.~\ref{fig:1}. 
A  semiconducting wire of length $L$ with SOI and in the ballistic regime is placed on the  surface of a bulk $s$-wave SC and we assume that the two are connected through a weakly transparent interface (tunneling regime). 
An external magnetic field is applied, oriented along the wire axis. 
We assume that orbital effects in the structure are small compared to the Zeeman splitting effect. Furthermore, the typical $g$ factor of electrons in the  wire is much larger than the one in the SC. 
In virtue of that, we  neglect the Zeeman term in the SC.

The order parameter $\Delta(\bm r)$ 
is generally non-uniform at the interface, and, typically, smaller than in the bulk  of the SC.
However, due to the low-transparency interface, we neglect the influence of the wire on $\Delta(\bm r)$  in the SC near the interface 
and use  the constant bulk value $\Delta$ within the usual BCS model.

The Hamiltonian of the SC-semiconducting wire  system, Fig.~\ref{fig:1}(a), can be written as $H= H_{\mathrm{BCS}}+H_{\mathrm{w}}+H_{\mathrm{t}}$.
In a three-dimensional clean SC one has $H_{\mathrm{BCS}} = (1/2)\int {\rm d}^3\rho~ \hat\Psi^{\dag}(\brho)\mathcal{H}_{\mathrm{BCS}}\hat\Psi(\brho)$, where
\begin{equation}
\mathcal{H}_{\mathrm{BCS}} = -(\bnabla^2/2m +\mu)\tau_z +\Delta\tau_x,
\end{equation}
$\hat\Psi(\brho) = (\hat\psi_{\uparrow}(\brho),\hat\psi_{\downarrow}(\brho), 
\hat\psi^{\dag}_{\downarrow}(\brho),-\hat\psi^{\dag}_{\uparrow}(\brho))^{\textrm{T}} $ is the electron operator in Nambu form in the SC, $m$ and  $\mu$
are the mass and the chemical potential. The Pauli matrices $\tau_i$ act in particle/hole space.
The coordinates are chosen such that the $\hat x$-axis is parallel to the wire and the $\hat y$-axis is perpendicular to the surface of SC. 
Space coordinates in the SC are denoted by $\brho=(\rho_x,\rho_y,\rho_z)$.

In the quasi-one dimensional wire  one has the Hamiltonian $H_{\mathrm{w}} = (1/2)\int {\rm d}x~\hat\Phi^{\dag}(x)\mathcal{H}_{\mathrm{w}}\hat\Phi(x)$, where
\begin{eqnarray}
\mathcal{H}_{\mathrm{w}} = -\left(\partial^2_x/2m_{\mathrm{w}}  +\mu_{\mathrm{w}} 
+i \alpha\sigma_z \partial_x \right)\tau_z
 + \Delta_{\rm Z}\sigma_x ,
\end{eqnarray}
$\hat\Phi(x) = (\phi_{\uparrow}(x),\phi_{\downarrow}(x), 
\phi^{\dag}_{\downarrow}(x),-\phi^{\dag}_{\uparrow}(x))^{\textrm{T}} $ is the electron operator in Nambu form in the wire; $\alpha$ is the spin-orbit 
coupling constant; $\Delta_{\rm Z}$ is the Zeeman energy; $m_{\mathrm{w}}$ and $\mu_{\mathrm{w}}$ are the effective mass and chemical potential in the wire. 
The Pauli matrices $\sigma_i$ act in spin space.

The tunneling between the SC and the wire is described by
\begin{equation}
H_{\mathrm{t}}=\frac{1}{2}\int {\rm d}x ~{\rm d}^3\rho~\left[ \hat\Psi^{\dag} (\brho) T_{\brho, x}  \hat\Phi(x) +{\rm H.c.}\right]\;,
\end{equation} 
where $T_{\brho, x} = t_{\brho, x}( 1+\tau_z)/2- t_{\brho, x}^*( 1-\tau_z)/2$ and
$t_{\brho,x}$ is the tunneling matrix element between the point $\brho$ in the SC and the point $x$ in the wire. 
We assume that tunneling occurs along the quasi one-dimensional interface, parallel to the $\hat x$-axis, in a  width of the order of the Fermi wavelength in the SC.

Integrating out the superconducting degrees of freedom and
including contributions proportional to the second order of the tunneling matrix element, we obtain the following equation for the Green function in the wire,
\begin{eqnarray}\label{g0}
g^{-1}(x)g(x,x' )-\int {\rm d}x_1\Sigma(x, x_1)g(x_1,x')
 = \delta(x-x')\;,\;\;\;\;
\end{eqnarray}
where 
\begin{equation}
g^{-1}(x)=\omega +\left(\partial^2_x/2m_{\mathrm{w}} +\mu_{\mathrm{w}} +i\alpha\sigma_z \partial_x \right)\tau_z -\Delta_{\rm Z}\sigma_x
\end{equation}
gives the zeroth-order wire Green function at energy $\omega$.
The second term on the left-hand side of Eq.~(\ref{g0}) represents 
the electron self-energy due to tunneling between wire and SC,
\begin{eqnarray}\label{selfE}
\Sigma(x, x_1) =  \int {\rm d}^3\rho ~{\rm d}^3\rho_1 ~
T_{x,\brho} ~G_{\omega}(\brho,\brho_1)~T^{*}_{\brho_1,x_1},
\end{eqnarray} 
and describes Andreev as well as normal elastic cotunneling processes.
The Green function of a clean homogeneous three-dimensional SC is given by
\begin{align}\label{GS}
G_{\omega}(\brho) = &-\frac{\pi \nu}{p_\mathrm{F}\rho} \bigg[  \tau_z\cos(p_\mathrm{F} \rho)  \\
&+ \frac{\omega + \Delta\tau_x}{\sqrt{\Delta^2-\omega^2}}\sin(p_\mathrm{F}\rho) \bigg] e^{ -\rho\sqrt{\Delta^2-\omega^2}/v_\mathrm{F}}, \nonumber
\end{align}
where $\nu = mp_\mathrm{F}/2\pi^2$ is the single-spin normal-state density of states, $\rho=|\brho|$, 
$v_{ \textrm{F}}$ and $p_{ \textrm{F}}=mv_{ \textrm{F}}$ are the Fermi velocity and momentum. 

We assume a uniform tunnel junction with local tunneling,  conserving electron spin and momentum component parallel to the wire. 
The real-space tunneling matrix element is given by
\begin{equation}\label{Tun}
t_{\brho,x} = t \delta(\rho_x-x) \delta(\rho_y)\delta(\rho_z),
\end{equation}
where we assume $t$ to be a real  constant.

We now proceed in two logical steps, similarly to what one does when finding first an isolated MF solution for a semi-infinite wire, and then using that result to deduce the finite-length splitting of two almost-decoupled MFs.
We first solve Eq.~(\ref{g0}) together with Eqs.~(\ref{selfE}) and (\ref{Tun}) in the $L\rightarrow\infty$ limit.
Taking into account that the Fermi momentum in the semiconducting wire is much smaller than that in the SC, one 
arrives at the equation
\begin{eqnarray}\label{g}
\left[g^{-1}(x) +\Gamma \frac{\omega-\Delta\tau_x}{\sqrt{\Delta^2-\omega^2}}\right]g(x,x' )  = \delta(x-x' ).
\end{eqnarray}
The parameter $\Gamma $ defines the properties of the tunneling interface and in this geometry is  given by 
\begin{equation}\label{Gamma}
\Gamma = \nu\pi^2t^2/2p_\mathrm{F}\;.
\end{equation}

The first term in square brackets in Eq.~(\ref{GS}), proportional to $\tau_z$, produces a renormalization of the chemical potential $\mu_{\rm w}$, corresponding to the doping effect of the SC, and can be absorbed by redefining it. In Eq.~(\ref{g}) and in the following it will be contained in $\mu_{\rm w}$.  %(additional gating in an experiment) 

One then sees that all the energies $\omega$ get renormalized, due to the second  term  in square brackets in Eq.~(\ref{GS}).
Finally, the proximity effect induces a mini-gap $\Delta_\star$ in the density of state of the wire, and at low frequencies $\omega\ll\Delta$ it given by~\cite{DasSarmaPRB2011,bib:PascalDisorder}
\begin{equation}\label{ProximityGap}
\Delta_\star= \Gamma\Delta/(\Gamma+\Delta) \;,
\end{equation}
approximated by $\Gamma$ in the tunneling regime $\Gamma\ll\Delta$, and by $\Delta$ in the transparent regime $\Gamma\gg\Delta$.
Even though we are formally working in the tunnel approximation, the results of this approach are known to be essentially correct up to $\Gamma\sim\Delta$~\cite{DasSarmaPRB2011}.

It is well known \cite{bib:AliceaReview} that Eq.~(\ref{g}) has a low-energy $p$-wave type SC phase when the Zeeman energy exceeds a critical value $\sqrt{\Delta_\star^2 + \mu^2_{\rm w}}$.
The upper limit of magnetic field which can be applied to the system is set by the
orbital effects that either destroy bulk superconductivity or induce orbital quantization in the wire that could suppress the proximity effect. 
In the $p-$wave phase, the system supports a MF bound state at each phase boundary.
In first approximation, a wire whose length is much larger than the localization length of the MF wave functions exhibits 
a doubly degenerate level with energy $E_{\rm \pm} =0$ pinned at the chemical potential, and
the two associated MF wave functions are approximately given by the solution of the semi-infinite problem, in the general form
$ {\Phi}_{\rm L}(x) = [u(x) ,v(x) , v^*(x), -u^*(x)]^\mathrm{T}$ and 
$ {\Phi}_\mathrm{R}(x)=i\sigma_x {\Phi}_\mathrm{{\rm L}}(L-x)$.

The exact form of the MF wave function is in general complicated. However, one can write down an approximate expression \cite{bib:Lena} 
in the weak-SOI regime $\Delta_{\rm Z} \gg \alpha^2 m_\mathrm{w}$ and far away from the topological transition, 
$\Delta_{\rm Z} - \Delta_\star\gg \sqrt{2\alpha^2 m_\mathrm{w}/\Delta_{\rm Z}} \Delta_\star$, assuming $\mu_{\rm w}=0$,
\begin{eqnarray}
 {\Phi}_{\rm L}(x) =  {\Phi}_{{\rm L}0}\sin(xp_{\mathrm{Fw}}) e^{-x/\xi_{\mathrm{w}}}/\sqrt{2\xi_\mathrm{w}}\;,
\end{eqnarray}
with spinor $ {\Phi}_{{\rm L}0} = (1,-1,-i,-i)^{\textrm{T}} e^{-i\pi/4} $. The wave function oscillates with wavevector 
$p_{\mathrm{F}\mathrm{w}} \simeq \sqrt{2m_\mathrm{w}\Delta_{\rm Z}}$ and decays on the scale of the localization length given by
\begin{equation}\label{xi_w}
\xi_{\mathrm{w}} \simeq \Delta_{\rm Z}/\alpha m_\mathrm{w}\Gamma 
\gg 1/\alpha m_{\rm w} \gg 1/p_{\mathrm{Fw}}\;.
\end{equation}
This weak-SOI regime, which is most relevant for current experiments~\cite{bib:Exp1, bib:Exp2, bib:Exp3, bib:Exp4, bib:Exp1Contr, bib:Exp2Contr}, allows us to obtain analytical results.
If $L\gtrsim\xi_{\rm w}$,  the energy splitting of the two MFs is exponentially suppressed, $E_{\rm +} \propto \exp(-L/\xi_\textrm{w})$. 
Similar considerations apply to the strong SOI limit, with the only difference that there are now two localization lengths associated 
with the $p\simeq 0$ and $p\simeq  p_{\mathrm{Fw}}$ regions of the spectrum \cite{bib:Lena}.

At low energies the electron operator $\hat\Phi(x)$ expansion in terms of eigen-excitations can be truncated and $\hat\Phi(x)$ is  projected onto the subspace of two decoupled 
MFs,
\begin{equation}\label{truncation}
\hat\Phi(x) =  {\Phi}_{\rm L}(x)\hat\gamma_{\rm L} +  {\Phi}_{\rm R}(x)\hat\gamma_{\rm R}\;,
\end{equation}
where the two MF operators $\hat\gamma_{ i} $ satisfy
$\hat\gamma_{ i} = \hat\gamma^{\dag}_{i}$ and $\{ \hat\gamma_i,\hat\gamma_j\}=\delta_{ij}$.
Within the same approximations, the electron Green function can be written as 
\begin{equation}\label{solution}
g(x,x' ) =  \left[{\Phi}_{\rm L}(x) {\Phi}^{*}_{\rm L}(x')\oplus 
                  {\Phi}_{\rm R}(x) {\Phi}^{*}_{\rm R}(x') \right]/2\omega.
\end{equation}

At this point we go back to Eq.~(\ref{g}) and introduce a finite length $L$ to the wire, and we analyze the resulting interaction between the two MF bound states as a perturbation to the result given in Eqs.~(\ref{truncation}) and (\ref{solution}), which refer to the $L\rightarrow\infty$ limit.

In order to obtain a quantitative estimate of the SC-mediated coupling of the two MFs, we evaluate the first-order correction $E_+$ to the MF state energy (see Supplementary material), determined by the expression (counted from $\mu_{\rm w}$)
\begin{equation}\label{E1}
E_+ =  t^2 \left\vert \int{\rm d}x~ {\rm d}x_1 \Phi^*_{\rm L}(x) \tau_zG_{\omega=0}(x, x_1)\tau_z \Phi_{\rm R}(x_1)\right\vert .
\end{equation}

We find that two MF edge states couple only via coherent tunneling through 
virtual quasiparticle states of the SC. This process is described by the first term in the square bracket in Eq.~(\ref{GS}), which in the infinite case only shifts the chemical potential.
Coupling through the anomalous Green function in the SC
is forbidden, because it would involve double-occupancy of the MF state by a split Cooper pair~\cite{bib:Zazunov}.

Assuming now that $L\gg \xi_\mathrm{w}$, and that therefore no appreciable direct overlap exists, we obtain for the SC-mediated energy splitting
\begin{eqnarray}\label{answer}
E_+ =\frac{2\Gamma}{\pi} \frac{p^2_{\mathrm{Fw}}}{p^3_\mathrm{F}\xi_\mathrm{w}}\frac{e^{-L/\xi_{\rm s}} }{p_\mathrm{F} L} |\cos(p_\mathrm{F}L)|\;.
\end{eqnarray}

{\it Discussion.---}% 
The prefactor $({p^2_{\mathrm{Fw}}}/{p^3_\mathrm{F}\xi_\mathrm{w}})(1/p_\mathrm{F} L)$ in Eq.~(\ref{answer}) originates from the ballistic motion in the SC, which provides a contribution $\propto (p_{\rm F} L)^{-1}$ to the energy splitting, 
as well as from the small ratio of number of conducting channels in the wire to the number of channels in the SC, which causes a further suppression factor $\propto p^2_\mathrm{Fw}/(p^3_\mathrm{F}\xi_\mathrm{w} )$. 
We observe here that this prefactor strongly depends on the geometry and the dimensionality of the contact between the wire and SC.
For example, in the case of point contact connection, the energy splitting of the MF state would be given by
$E_+\propto \Gamma_{\rm pc}\exp(-L/\xi_{\rm s}) \cos(p_\mathrm{F}L)/(p_{\rm F} L)$, with a prefactor $\Gamma_{\rm pc}$ defined by the point contact transparency.

Let us also comment on the dependence of the prefactor in Eq.~(\ref{answer}) on the dimensionality of the SC.
In the case of a wire coupled to ${d}$-dimensional ballistic SC, the prefactor originating from the
SC Green function scales as $\propto(p_{\mathrm{F}}L)^{(1-d)/2}$. The coupling between MFs is thus noticeably enhanced in lower dimensions. 
Alternatively, the prefactor can be increased in the limit of a diffusive SC, where disorder enhances the correlation effects as compared to a clean system.
In SCs with linear dimensions larger than the mean free path $\ell$,
the prefactor has the form $\propto(p^2_{\mathrm{F}}L\ell)^{(1-d)/4}$. 
The coherence length of the disordered SC is accordingly modified and becomes $\propto \sqrt{\xi_{\rm s} \ell}$.

The short-wire limit studied above is defined by $ \xi_{\rm s} \gtrsim L \gg \xi_\textrm{w}$, which translates into the following condition for the wire parameters [see Eq.~(\ref{xi_w})]:
\begin{equation}\label{short_wire_regime}
\frac{\Gamma}{\Delta} \gtrsim \frac{\Gamma}{E_{\rm T}} \gg \frac{\alpha}{v_\mathrm{F}}\frac{\Delta_{\rm Z}}{\alpha^2 m_\mathrm{w}}\;,
\end{equation}
where $E_{\rm T}=v_{\rm F}/L$ is the ballistic Thouless energy, associated  with the free motion over a scale $L$ in the SC.  %, and $v_{\rm so}=\alpha$ is the spin-orbit velocity. 
By taking for instance as typical values $\Delta$=1 meV, $\Gamma=0.25\Delta$, $L=1~\mu$m, $v_{\rm F}=10^6$ m/s, $\Delta_{\rm Z}=2\Gamma$, and $\alpha^2m_{\rm w}=50~\mu$eV$ = 0.1\Delta_{\rm Z}$, one satisfies the inequalities of Eq.~(\ref{short_wire_regime}). The large factor $(\Delta_{\rm Z}/\alpha^2m_{\rm w})$ is more than compensated by the difference between Fermi velocity in the SC and SOI velocity in the wire ($\simeq 3\cdot10^4$ m/s with the above parameters).

\begin{figure}[t]  \centering
\includegraphics[width=1.0\linewidth]{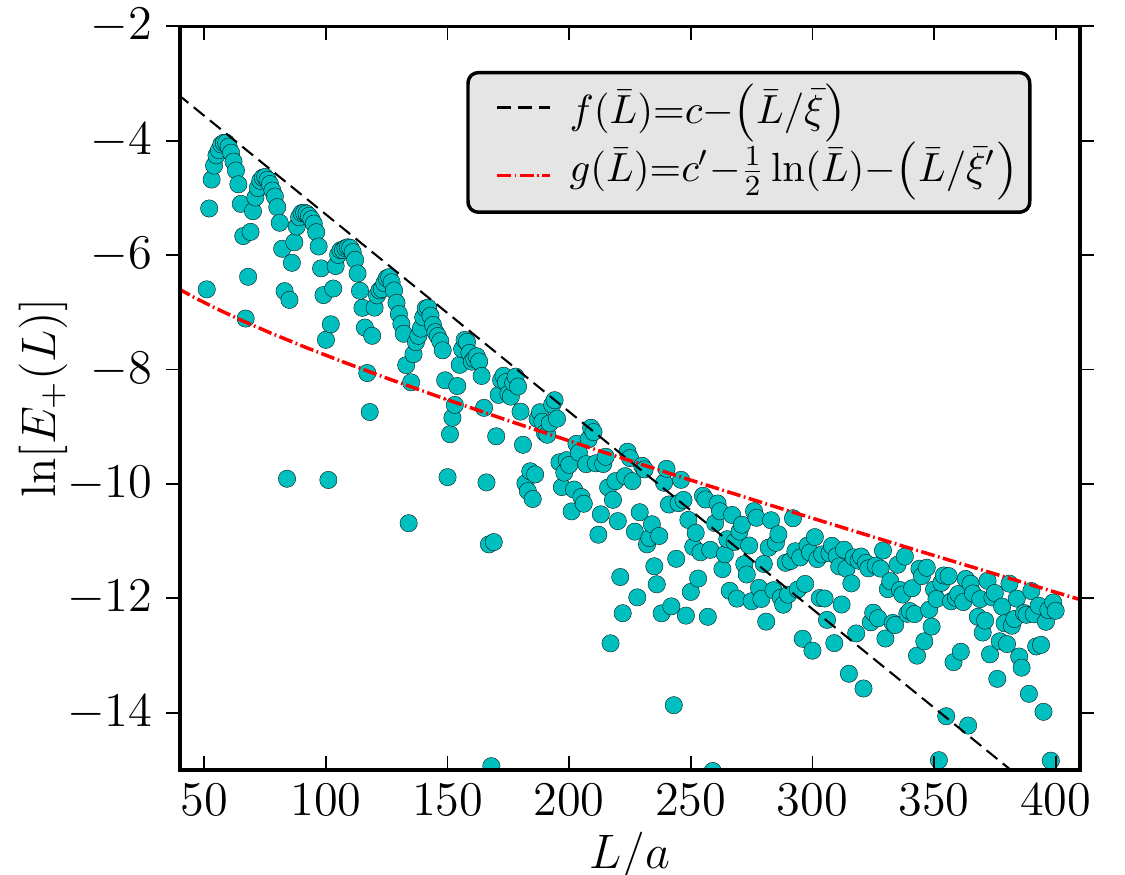}
\caption{%(a) Cartoon of the adopted tight-binding (TB) model. (b) 
Numerical results (cyan circles) for the energy splitting $E_+$ of the MFs as a function of wire length $L$, on (natural) logarithmic scale. 
The points do not lie on a straight line, showing that there is no pure exponential decay of the splitting.
The dashed black line describes a bare exponential behavior, as the wire-only model would predict, with associated decay length $\bar\xi=\xi_\mathrm{w}/a=25$ ($a$ tight-binding lattice constant).
The red dash-dotted curve refers to the prediction Eq.~(\ref{answer}) for a two-dimensional SC, $E_{+}(L)\simeq L^{-1/2}\exp({-L/\xi_{\rm s}})$, with $\bar\xi^\prime=\xi_{\rm s}/a=100$. 
Here we defined dimensionless quantities $\bar L=L/a$ and $\bar\xi=\xi/a$, with $a$ the lattice constant of the TB model. The constants $c, c^\prime,\xi$ and $\xi^\prime$ are free fit parameters.}
\label{fig:2}
\end{figure}

{\it Numerical simulations.---}We confirmed the existence of the SC-induced MF energy splitting, and in general the presence of two independent length scales, $\xi_{\rm w}$ and $\xi_{\rm s}$, by performing exact-diagonalization calculations of a two-dimensional tight-binding model for the wire-SC coupled system, investigating  the MF splitting as a function of $L$. 
In particular, we consider a quasi-1D wire of width $W_{\rm w}$ laterally coupled to a large SC ($W_{\rm s}\gg W_{\rm w}$) along all its length $L$.
The wire is described by a hopping parameter $t_{\rm w}$, finite SOI hopping $\alpha$ and Zeeman splitting $\Delta_z$. 
The SC has a larger  hopping energy $t_{\rm s}\gg t_{\rm w}$ and a uniform pairing potential $\Delta$, but zero SOI and Zeeman coupling. 
The wire-SC coupling is  modeled by an interface hopping $t$. 
For further details on the tight-binding model see Ref.~\cite{bib:Diego1}.
In Fig.~\ref{fig:2} we plot the numerical results for the MF energy as a function of $L$. One immediately notices that the data (their envelope) do not lie on a straight line, like they do in the wire-only case, and that therefore no pure exponential behavior is observed. 
Rather, the obtained data seem to be composed  of two contributions: 
one purely exponential, emerging at low values of $L$, with a characteristic decay length $\xi\simeq25a$ (black dashed line) and with wide oscillations visible up to $L\simeq200a$; we attribute this contribution to the direct-overlap splitting in the wire.
The other contribution, emerging at large $L$, follows a modified exponential behavior, with localization length $\xi^\prime=100a$ extracted at the largest $L$'s, and much faster oscillations. We attribute this contribution to the SC-mediated MF correlations.
The corresponding points can be successfully fitted by a $\sim \exp(-L/\xi^\prime)/\sqrt{L}$ curve (red dash-dotted line), which is the analytical prediction for this quasi-2D geometry [see Eq.~(\ref{answer}) and ensuing comments on the role of the SC dimensionality]. 
On such curve lie both the maxima at large $L$ and some of the relative minima at low $L$, where the SC contribution is secondary and can be observed in correspondence of the minima of the wire contribution.

{\it Coupling between two different wires.---}%
Let us finally consider another setup, see Fig.~\ref{fig:1}(b), 
where two different wires are placed close to each other on the surface of a bulk $s$-wave SC. 
At each wire/SC interface the contact is as before.
We further assume that the length $L$ of each wire is much longer that the wire separation $D$, and that $L\gg\xi_{\rm s},\xi_{\rm w}$. If $D\lesssim\xi_{\rm s}$, then the  same considerations made above apply and the same  result for the inter-wire splitting of the MFs is obtained. 
Since the proposed schemes for MF-based quantum computation involve multiple wires hosting MFs at their ends~\cite{bib:AliceaReview}, brought into vicinity to each other in order to perform logical braiding operations, one must be concerned about both {\it intra-wire} and {\it inter-wire} MF energy splittings.
In the two-wire case there is no direct overlap between the states $\Phi_{\rm R1}$ and $\Phi_{\rm L2}$ of Fig.~\ref{fig:1}(b), so that the SC-mediated splitting is the only present and is easier to be identified.

The coupling between two MFs from different wires can be tuned by changing the ``type'' (i.e. the phase) of the MFs that are interacting (real-part or imaginary-part type).
This can be accomplished for instance in a wire with N-S setup by varying the length of the N part~\cite{bib:Lena} or by changing the relative wire orientation~\cite{bib:Alicea_Nphys}.
The general expression for the inter-wire MF energy splitting is given in the Supplementary material. In the special case of two identical wires and two MFs of opposite type one obtains 
Eq.~(\ref{answer}), with $L$ replaced by $D$. In the case of two MF of the same type one has $E_+=0$.

{\it Conclusions.---}We have studied the coupling between two MFs arising in a wire with SOI placed in contact with an $s$-wave SC.
We have shown that if the separation between the two MFs is of the order of the SC coherence length, elastic coherent tunneling processes 
through virtual quasiparticles states in the SC split the doubly degenerate 
energy level of the MFs. This remains true even if the wire is much longer than the wire MF localization length, and the direct-overlap splitting is thus negligible.

\emph{Acknowledgments.---}We acknowledge helpful discussions with Luka Trifunovic, and support by the Swiss NSF, NCCR Nanoscience, NCCR QSIT, and the EU project SOLID.

\appendix
\section{Detailed Calculations for the energy splitting}
%\emph{Appendix}.-- 
The Hamiltonian of the topological superconducting wire is given by
\begin{equation}\label{MainEq}
H = \frac{1}{2}\int \mathrm{d}x \mathrm{d}x' \hat{\Phi}^{\dag}(x) \mathcal{H}(x,x')\hat{\Phi}(x') \;,
\end{equation}
where 
\begin{equation}
\mathcal{H}(x,x') = \delta(x-x')g^{-1}(x) - t^2 \tau_z G_\omega(x,x')\tau_z.
\end{equation}
The operator is written in Nambu form $\hat{\Phi}(x)=(\hat{\Phi}_{j\uparrow}(x), \hat{\Phi}_{j\downarrow}(x), \hat{\Phi}^{\dag}_{j\downarrow}(x), -\hat{\Phi}^{\dag}_{j\uparrow}(x))^{\mathrm{T}} $.
The Bogoliubov-de Gennes equation for an eigenstate $\Phi_{\omega}(x)$ is written as
\begin{align}\label{Appendix-Ham1}\nonumber
g^{-1}(x)\Phi_{\omega}(x) -t^2 \int_{0}^{L} \mathrm{d}x' \tau_z G_{\omega}(x,x')\tau_z &\Phi_{\omega}(x') \\
=\omega &\Phi_{\omega}(x) ,
\end{align}
and the corresponding operator reads
\begin{equation}
\hat{\gamma}_{\omega} = \int \mathrm{d}x ~\Phi_{\omega}^*(x) \hat{\Phi}(x),
\end{equation}
in terms of which the Hamiltonian Eq.~(\ref{MainEq}) takes the diagonal form
\begin{equation}
H = \frac{1}{2}\sum_{\omega} \omega \hat{\gamma}^{\dag}_{\omega} \hat{\gamma}_{\omega} .
\end{equation}
The Green function in Eq. (\ref{Appendix-Ham1}) in the homogeneous case is given by
\begin{align}\label{GS2}
G_{\omega}(\mathbf{ r}) = &-\frac{\pi \nu}{p_\mathrm{F}r} \bigg[  \tau_z\cos(p_\mathrm{F} r)  \\
&+ \frac{\omega + \Delta\tau_x}{\sqrt{\Delta^2-\omega^2}}\sin(p_\mathrm{F}r) \bigg] e^{ -r\sqrt{\Delta^2-\omega^2}/v_\mathrm{F}}\;, \nonumber
\end{align}
where $r=|x-x'|$. The term $\propto \tau_x$ yields the anomalous Green function $F(r)$ and the rest gives the single-particle Green function.

In first-order approximation we assume that $L\gg \xi_{\rm s} =\Delta/v_\mathrm{F}\gg\xi_{\mathrm{w}}$ and solve Eq.~(\ref{Appendix-Ham1}) separately in the vicinity of 
$x\sim 0$ and of $x\sim L$, {\it i.e.} left and right edges of the wire. In the vicinity of the left edge of the wire ($x\simeq0$)
\begin{align}\label{Appendix-Ham12}
g^{-1}(x)\Phi_{\omega}(x) -t^2 \int_{0}^{L} \mathrm{d}x' \tau_zG_{\omega}(x,x')\tau_z\Phi_{\omega}(x') =\omega \Phi_{\omega}(x)  \;.
\end{align}
In the vicinity of the right edge ($x\simeq L$) one has an identical equation, but with $x$ replaced by $(L-x)$.

Analogous considerations can be applied to the case of two wires of length $L$ placed along the $\hat x$ direction and 
separated by a distance $D$. One has in the left wire ($-L<x<0$) the equation
\begin{align}\label{Appendix-Ham2}\nonumber
g_1^{-1}(x)\Phi_{\omega}(x) -t^2 \int_{-L}^{L+D} \mathrm{d}x' \tau_zG_{\omega}(x,x')\tau_z &\Phi_{\omega}(x') \\
=\omega &\Phi_{\omega}(x) 
\end{align}
and in the right wire ($D<x<D+L$)
\begin{align}\label{Appendix-Ham3}\nonumber
g_2^{-1}(x)\Phi_{\omega}(x) -t^2 \int_{-L}^{D+L} \mathrm{d}x' \tau_zG_{\omega}(x,x')\tau_z&\Phi_{\omega}(x') \\
=\omega &\Phi_{\omega}(x) \;.
\end{align}
Here $g_{1,2}^{-1}(x)$ are defined for two wires that can have in general different Zeeman terms, SOI strengths, etc. 

Let us go back now to the single wire case. 
We first solve the problem in the vicinity of the left edge, $x\sim 0$,
\begin{equation}\label{Appendix-Ham}
g^{-1}(x)\Phi_{\omega}(x) +\Gamma\frac{\omega-\Delta\tau_x}{\sqrt{\Delta^2-\omega^2}}\Phi_{\omega}(x) =\omega \Phi_{\omega}(x)\;.
\end{equation}
One gets a MF solution corresponding to energy $\omega=0$.
The two MF wave functions (at the left and right edges of the wire) can be written as (in what follows we omit the index $\omega=0$)
\begin{align}\label{Phi_j}
\Phi_{j}(x) = (u_{j}(x),v_{j}(x),v_{j}^*(x),-u_{j}^*(x))^{\mathrm{T}}, \quad j={\rm L, {\rm R}} .
\end{align}
One can write solutions at the left and right edges of the wire in the limit of large magnetic field as
\begin{align}
\Phi_{\rm L}(x) &= (1,-1,-i,-i)^{\mathrm{T}}e^{-i\pi/4} \frac{\sin(xp_{\mathrm{Fw}})}{\sqrt{2\xi_\mathrm{w}}}e^{-x/\xi_\mathrm{w}}\;,\\ \nonumber
\Phi_{\rm R}(x) &= (-1,1,-i,-i)^{\mathrm{T}}e^{i\pi/4}\frac{\sin[(x-L)p_{\mathrm{Fw}}]}{\sqrt{2\xi_\mathrm{w}}}e^{(x-L)/\xi_\mathrm{w}}\;.
\end{align}
We note that the two solutions, beyond satisfying the particle-hole symmetry, are related by   
\begin{equation}
\Phi_{\rm R}(L-x) = i\sigma_x\Phi_{\rm L}(x) \;,
\end{equation}
owing to the symmetry of the single wire Hamiltonian Eq.~(\ref{Appendix-Ham}).
The hybridization of the zero-energy levels due to the finite size of the wire is similar 
to hybridization in a double-quantum-dot system. The degenerate zero energy level splits into two levels $E_{\pm}$ with corresponding wave functions
that can be written as 
\begin{equation}
\Phi_{\pm}(x) = \frac{1}{\sqrt 2}\bigg[\Phi_{{\rm L}}(x) \pm e^{i\varphi}\Phi_{{\rm R}}(x)\bigg] .
\end{equation} 
The energy levels are related by the particle-hole symmetry $E_{+}=-E_{-}$. The same particle-hole symmetry of the Hamiltonian (\ref{MainEq}) requires 
\begin{equation}\Phi_{+}(x)=\tau_y\sigma_y\Phi^*_{-}(x) .
\end{equation}
One finds that $\varphi = \pi/2$.
The half-energy splitting of two MFs is given by the overlap of two wave functions $\Psi_{j}(x)$ localized at the edges the wire.
Within degenerate perturbation theory we obtain for the single-wire case
\begin{align}\nonumber
E^2_{+} = t^4 &\bigg[\int_{0}^{L/2} \mathrm{d}x \int_{L/2}^{L} \mathrm{d}x' \Phi^{*}_{{\rm L}}(x)\tau_zG_{\omega=0}(x,x')\tau_z\Phi_{{\rm R}}(x')\bigg]\\
\times &\bigg[\int_{0}^{L/2} \mathrm{d}x\int_{L/2}^{L} \mathrm{d}x' \Phi^{*}_{{\rm R}}(x')\tau_zG_{\omega=0}(x',x)\tau_z\Phi_{{\rm L}}(x) \bigg],
\end{align}
and for two wires
\begin{align}\nonumber
E^2_{+} = t^4 &\int_{-L}^{0} \mathrm{d}x \int_{D}^{L+D} \mathrm{d}x' \Phi^{*}_{{\rm R}1}(x)\tau_zG_{\omega=0}(x,x')\tau_z\Phi_{{\rm L}2}(x')\\ \times
&\int_{D}^{D+L} \mathrm{d}x \int_{-L}^{0}\mathrm{d}x' \Phi^{*}_{{\rm L}2}(x)\tau_zG_{\omega=0}(x,x')\tau_z\Phi_{{\rm R}1}(x').
\end{align}
The indexes ${\rm R}1$ and ${\rm L}2$ identify the solutions of Eqs.~(\ref{Appendix-Ham2}), (\ref{Appendix-Ham3}) at the right ($x\sim 0$)  and left  
($x\sim D$) parts of two wires, respectively. Expressing the Green function of the superconductor as (see Eq. \ref{GS2}) 
\begin{equation}
G_{\omega=0}(x,x') =G(x-x')\tau_z+F(x-x')\tau_x.
\end{equation}
In the single-wire case the term $\Phi^{*}_{\rm L}(x) \tau_x\Phi_{\rm R}(x)$ vanishes and we are left with the diagonal term $\Phi^{*}_{\rm L}(x) \tau_z\Phi_{\rm R}(x) \neq 0$ only.
We obtain
\begin{align}\nonumber
E_{+}&=\frac{2\pi \nu t^2}{p_\mathrm{F}\xi_{\mathrm{w}}}\Bigg|\int_{0}^{L/2} \mathrm{d}x \int_{L/2}^{L}\mathrm{d}x' \frac{e^{-|x-x'|/\xi_{\rm s}}}{|x-x'|} e^{-(x+L-x')/\xi_\mathrm{w}}\\ 
&\times \cos\left[p_F(x-x')\right]\sin(xp_{\mathrm{Fw}})\sin\left[(x'-L)p_{\mathrm{Fw}}\right]\Bigg|.
\end{align}
We can replace $|x-x'|\rightarrow L$ except in the fast oscillating term $\cos\left[(x-x')p_\mathrm{F}\right]$. 

One obtains in the limit $L\gtrsim \xi_{\rm s} \gg\xi_\mathrm{w}$
\begin{align}
E_{+}&=\frac{2\pi \nu t^2e^{-L/\xi_{\rm s}}}{p_\mathrm{F}L\xi_{\mathrm{w}}}\Bigg|\int_{0}^{L} \mathrm{d}x\mathrm{d}x'  \\ \nonumber
&\times \cos[(x-x')p_{\rm F}]\sin(xp_{\mathrm{Fw}})\sin[(x'-L)p_{\mathrm{Fw}}]\Bigg|.
\end{align}
The energy splitting for the single wire is given as
\begin{eqnarray}
E_+ =\frac{2\Gamma}{\pi} \frac{p^2_{\mathrm{Fw}}}{p^3_\mathrm{F}\xi_\mathrm{w}}\frac{e^{-L/\xi_{\rm s}} }{p_\mathrm{F} L} |\cos(p_\mathrm{F}L)|.
\end{eqnarray}
Using Eq.~(\ref{Phi_j}), for the two-wire case we obtain
\begin{align}\label{TwoW}
E_{+}=  2t^2 &\Bigg|\int_{-L}^{0} \mathrm{d}x \int_{D}^{L+D}\mathrm{d}x' \\ \nonumber
&\times\bigg\{ F(x-x') \mathrm{Im}\left[v_{\rm L}(x)u_{\rm R}(x')  -u_{\rm L}(x)v_{\rm R}(x')\right] \\\nonumber
&+ G(x-x') \mathrm{Im}\left[u^*_{\rm L}(x)u_{\rm R}(x')+  v^*_{\rm L}(x)v_{\rm R}(x')\right]
\bigg\}\Bigg| .
\end{align}
The energy splitting generally depends on the explicit form of the MF wave functions in the two wires.

\bibliography{CouplingMF}

\begin{thebibliography}{39}
\expandafter\ifx\csname natexlab\endcsname\relax\def\natexlab#1{#1}\fi
\expandafter\ifx\csname bibnamefont\endcsname\relax
  \def\bibnamefont#1{#1}\fi
\expandafter\ifx\csname bibfnamefont\endcsname\relax
  \def\bibfnamefont#1{#1}\fi
\expandafter\ifx\csname citenamefont\endcsname\relax
  \def\citenamefont#1{#1}\fi
\expandafter\ifx\csname url\endcsname\relax
  \def\url#1{\texttt{#1}}\fi
\expandafter\ifx\csname urlprefix\endcsname\relax\def\urlprefix{URL }\fi
\providecommand{\bibinfo}[2]{#2}
\providecommand{\eprint}[2][]{\url{#2}}

\bibitem[{\citenamefont{Wilczek}(2009)}]{bib:WilczekReview}
\bibinfo{author}{\bibfnamefont{F.}~\bibnamefont{Wilczek}},
  \bibinfo{journal}{Nat. Phys.} \textbf{\bibinfo{volume}{5}},
  \bibinfo{pages}{614} (\bibinfo{year}{2009}).

\bibitem[{\citenamefont{Alicea}(2012)}]{bib:AliceaReview}
\bibinfo{author}{\bibfnamefont{J.}~\bibnamefont{Alicea}},
  \bibinfo{journal}{Reports on Progress in Physics}
  \textbf{\bibinfo{volume}{75}}, \bibinfo{pages}{076501}
  (\bibinfo{year}{2012}).

\bibitem[{\citenamefont{Franz}(2013)}]{bib:FranzReview}
\bibinfo{author}{\bibfnamefont{M.}~\bibnamefont{Franz}}, \bibinfo{journal}{Nat.
  Nanotech.} \textbf{\bibinfo{volume}{8}}, \bibinfo{pages}{149}
  (\bibinfo{year}{2013}).

\bibitem[{\citenamefont{Mourik et~al.}(2012)\citenamefont{Mourik, Zuo, Frolov,
  Plissard, Bakkers, and Kouwenhoven}}]{bib:Exp1}
\bibinfo{author}{\bibfnamefont{V.}~\bibnamefont{Mourik}},
  \bibinfo{author}{\bibfnamefont{K.}~\bibnamefont{Zuo}},
  \bibinfo{author}{\bibfnamefont{S.~M.} \bibnamefont{Frolov}},
  \bibinfo{author}{\bibfnamefont{S.~R.} \bibnamefont{Plissard}},
  \bibinfo{author}{\bibfnamefont{E.~P. A.~M.} \bibnamefont{Bakkers}},
  \bibnamefont{and} \bibinfo{author}{\bibfnamefont{L.~P.}
  \bibnamefont{Kouwenhoven}}, \bibinfo{journal}{Science}
  \textbf{\bibinfo{volume}{336}}, \bibinfo{pages}{1003} (\bibinfo{year}{2012}).

\bibitem[{\citenamefont{Deng et~al.}(2012)\citenamefont{Deng, Yu, Huang,
  Larsson, Caroff, and Xu}}]{bib:Exp2}
\bibinfo{author}{\bibfnamefont{M.~T.} \bibnamefont{Deng}},
  \bibinfo{author}{\bibfnamefont{C.~L.} \bibnamefont{Yu}},
  \bibinfo{author}{\bibfnamefont{G.~Y.} \bibnamefont{Huang}},
  \bibinfo{author}{\bibfnamefont{M.}~\bibnamefont{Larsson}},
  \bibinfo{author}{\bibfnamefont{P.}~\bibnamefont{Caroff}}, \bibnamefont{and}
  \bibinfo{author}{\bibfnamefont{H.~Q.} \bibnamefont{Xu}},
  \bibinfo{journal}{Nano Letters} \textbf{\bibinfo{volume}{12}},
  \bibinfo{pages}{6414} (\bibinfo{year}{2012}).

\bibitem[{\citenamefont{Das et~al.}(2012)\citenamefont{Das, Ronen, Most, Oreg,
  Heiblum, and Shtrikman}}]{bib:Exp3}
\bibinfo{author}{\bibfnamefont{A.}~\bibnamefont{Das}},
  \bibinfo{author}{\bibfnamefont{Y.}~\bibnamefont{Ronen}},
  \bibinfo{author}{\bibfnamefont{Y.}~\bibnamefont{Most}},
  \bibinfo{author}{\bibfnamefont{Y.}~\bibnamefont{Oreg}},
  \bibinfo{author}{\bibfnamefont{M.}~\bibnamefont{Heiblum}}, \bibnamefont{and}
  \bibinfo{author}{\bibfnamefont{H.}~\bibnamefont{Shtrikman}},
  \bibinfo{journal}{Nat Phys} \textbf{\bibinfo{volume}{8}},
  \bibinfo{pages}{887} (\bibinfo{year}{2012}).

\bibitem[{\citenamefont{Rokhinson et~al.}(2012)\citenamefont{Rokhinson, Liu,
  and Furdyna}}]{bib:fractionalJ}
\bibinfo{author}{\bibfnamefont{L.~P.} \bibnamefont{Rokhinson}},
  \bibinfo{author}{\bibfnamefont{X.}~\bibnamefont{Liu}}, \bibnamefont{and}
  \bibinfo{author}{\bibfnamefont{J.~K.} \bibnamefont{Furdyna}},
  \bibinfo{journal}{Nat Phys} \textbf{\bibinfo{volume}{8}},
  \bibinfo{pages}{795} (\bibinfo{year}{2012}).

\bibitem[{\citenamefont{Williams et~al.}(2012)\citenamefont{Williams, Bestwick,
  Gallagher, Hong, Cui, Bleich, Analytis, Fisher, and
  Goldhaber-Gordon}}]{bib:unconventionalJ}
\bibinfo{author}{\bibfnamefont{J.~R.} \bibnamefont{Williams}},
  \bibinfo{author}{\bibfnamefont{A.~J.} \bibnamefont{Bestwick}},
  \bibinfo{author}{\bibfnamefont{P.}~\bibnamefont{Gallagher}},
  \bibinfo{author}{\bibfnamefont{S.~S.} \bibnamefont{Hong}},
  \bibinfo{author}{\bibfnamefont{Y.}~\bibnamefont{Cui}},
  \bibinfo{author}{\bibfnamefont{A.~S.} \bibnamefont{Bleich}},
  \bibinfo{author}{\bibfnamefont{J.~G.} \bibnamefont{Analytis}},
  \bibinfo{author}{\bibfnamefont{I.~R.} \bibnamefont{Fisher}},
  \bibnamefont{and}
  \bibinfo{author}{\bibfnamefont{D.}~\bibnamefont{Goldhaber-Gordon}},
  \bibinfo{journal}{Phys. Rev. Lett.} \textbf{\bibinfo{volume}{109}},
  \bibinfo{pages}{056803} (\bibinfo{year}{2012}).

\bibitem[{\citenamefont{Churchill et~al.}(2013)\citenamefont{Churchill, Fatemi,
  Grove-Rasmussen, Deng, Caroff, Xu, and Marcus}}]{bib:Exp4}
\bibinfo{author}{\bibfnamefont{H.~O.~H.} \bibnamefont{Churchill}},
  \bibinfo{author}{\bibfnamefont{V.}~\bibnamefont{Fatemi}},
  \bibinfo{author}{\bibfnamefont{K.}~\bibnamefont{Grove-Rasmussen}},
  \bibinfo{author}{\bibfnamefont{M.}~\bibnamefont{Deng}},
  \bibinfo{author}{\bibfnamefont{P.}~\bibnamefont{Caroff}},
  \bibinfo{author}{\bibfnamefont{H.~Q.} \bibnamefont{Xu}}, \bibnamefont{and}
  \bibinfo{author}{\bibfnamefont{C.~M.} \bibnamefont{Marcus}}
  (\bibinfo{year}{2013}), \bibinfo{note}{arXiv:1303.2407}.

\bibitem[{\citenamefont{Lee et~al.}(2012)\citenamefont{Lee, Jiang, Aguado,
  Katsaros, Lieber, and De~Franceschi}}]{bib:Exp1Contr}
\bibinfo{author}{\bibfnamefont{E.~J.~H.} \bibnamefont{Lee}},
  \bibinfo{author}{\bibfnamefont{X.}~\bibnamefont{Jiang}},
  \bibinfo{author}{\bibfnamefont{R.}~\bibnamefont{Aguado}},
  \bibinfo{author}{\bibfnamefont{G.}~\bibnamefont{Katsaros}},
  \bibinfo{author}{\bibfnamefont{C.~M.} \bibnamefont{Lieber}},
  \bibnamefont{and}
  \bibinfo{author}{\bibfnamefont{S.}~\bibnamefont{De~Franceschi}},
  \bibinfo{journal}{Phys. Rev. Lett.} \textbf{\bibinfo{volume}{109}},
  \bibinfo{pages}{186802} (\bibinfo{year}{2012}).

\bibitem[{\citenamefont{Lee et~al.}(2013)\citenamefont{Lee, Jiang, Houzet,
  Aguado, Lieber, and De~Franceschi}}]{bib:Exp2Contr}
\bibinfo{author}{\bibfnamefont{E.~J.~H.} \bibnamefont{Lee}},
  \bibinfo{author}{\bibfnamefont{X.}~\bibnamefont{Jiang}},
  \bibinfo{author}{\bibfnamefont{M.}~\bibnamefont{Houzet}},
  \bibinfo{author}{\bibfnamefont{R.}~\bibnamefont{Aguado}},
  \bibinfo{author}{\bibfnamefont{C.~M.} \bibnamefont{Lieber}},
  \bibnamefont{and}
  \bibinfo{author}{\bibfnamefont{S.}~\bibnamefont{De~Franceschi}}
  (\bibinfo{year}{2013}), \bibinfo{note}{arXiv:1302.2611}.

\bibitem[{\citenamefont{Volovik}(2009)}]{bib:Volovik1}
\bibinfo{author}{\bibfnamefont{G.}~\bibnamefont{Volovik}},
  \bibinfo{journal}{JETP Lett.} \textbf{\bibinfo{volume}{90}},
  \bibinfo{pages}{440} (\bibinfo{year}{2009}).

\bibitem[{\citenamefont{Volovik}(2003)}]{bib:Volovik-Review}
\bibinfo{author}{\bibfnamefont{G.}~\bibnamefont{Volovik}},
  \emph{\bibinfo{title}{The Universe in a Helium Droplet}}
  (\bibinfo{publisher}{Oxford University Press, Oxford}, \bibinfo{year}{2003}).

\bibitem[{\citenamefont{Read and Green}(2000)}]{bib:Read}
\bibinfo{author}{\bibfnamefont{N.}~\bibnamefont{Read}} \bibnamefont{and}
  \bibinfo{author}{\bibfnamefont{D.}~\bibnamefont{Green}},
  \bibinfo{journal}{Phys. Rev. B} \textbf{\bibinfo{volume}{61}},
  \bibinfo{pages}{10267} (\bibinfo{year}{2000}).

\bibitem[{\citenamefont{Nayak et~al.}(2008)\citenamefont{Nayak, Simon, Stern,
  Freedman, and Das~Sarma}}]{bib:Nayak}
\bibinfo{author}{\bibfnamefont{C.}~\bibnamefont{Nayak}},
  \bibinfo{author}{\bibfnamefont{S.~H.} \bibnamefont{Simon}},
  \bibinfo{author}{\bibfnamefont{A.}~\bibnamefont{Stern}},
  \bibinfo{author}{\bibfnamefont{M.}~\bibnamefont{Freedman}}, \bibnamefont{and}
  \bibinfo{author}{\bibfnamefont{S.}~\bibnamefont{Das~Sarma}},
  \bibinfo{journal}{Rev. Mod. Phys.} \textbf{\bibinfo{volume}{80}},
  \bibinfo{pages}{1083} (\bibinfo{year}{2008}).

\bibitem[{\citenamefont{Salomaa and Volovik}(1988)}]{bib:Salomaa}
\bibinfo{author}{\bibfnamefont{M.~M.} \bibnamefont{Salomaa}} \bibnamefont{and}
  \bibinfo{author}{\bibfnamefont{G.~E.} \bibnamefont{Volovik}},
  \bibinfo{journal}{Phys. Rev. B} \textbf{\bibinfo{volume}{37}},
  \bibinfo{pages}{9298} (\bibinfo{year}{1988}).

\bibitem[{\citenamefont{Chung and Zhang}(2009)}]{bib:Zhang}
\bibinfo{author}{\bibfnamefont{S.~B.} \bibnamefont{Chung}} \bibnamefont{and}
  \bibinfo{author}{\bibfnamefont{S.-C.} \bibnamefont{Zhang}},
  \bibinfo{journal}{Phys. Rev. Lett.} \textbf{\bibinfo{volume}{103}},
  \bibinfo{pages}{235301} (\bibinfo{year}{2009}).

\bibitem[{\citenamefont{Fu and Kane}(2008)}]{bib:Fu}
\bibinfo{author}{\bibfnamefont{L.}~\bibnamefont{Fu}} \bibnamefont{and}
  \bibinfo{author}{\bibfnamefont{C.~L.} \bibnamefont{Kane}},
  \bibinfo{journal}{Phys. Rev. Lett.} \textbf{\bibinfo{volume}{100}},
  \bibinfo{pages}{096407} (\bibinfo{year}{2008}).

\bibitem[{\citenamefont{Tanaka et~al.}(2009)\citenamefont{Tanaka, Yokoyama, and
  Nagaosa}}]{bib:Tanaka}
\bibinfo{author}{\bibfnamefont{Y.}~\bibnamefont{Tanaka}},
  \bibinfo{author}{\bibfnamefont{T.}~\bibnamefont{Yokoyama}}, \bibnamefont{and}
  \bibinfo{author}{\bibfnamefont{N.}~\bibnamefont{Nagaosa}},
  \bibinfo{journal}{Phys. Rev. Lett.} \textbf{\bibinfo{volume}{103}},
  \bibinfo{pages}{107002} (\bibinfo{year}{2009}).

\bibitem[{\citenamefont{Sato and Fujimoto}(2009)}]{bib:Sato}
\bibinfo{author}{\bibfnamefont{M.}~\bibnamefont{Sato}} \bibnamefont{and}
  \bibinfo{author}{\bibfnamefont{S.}~\bibnamefont{Fujimoto}},
  \bibinfo{journal}{Phys. Rev. B} \textbf{\bibinfo{volume}{79}},
  \bibinfo{pages}{094504} (\bibinfo{year}{2009}).

\bibitem[{\citenamefont{Lutchyn et~al.}(2010)\citenamefont{Lutchyn, Sau, and
  Das~Sarma}}]{bib:Lutchyn}
\bibinfo{author}{\bibfnamefont{R.~M.} \bibnamefont{Lutchyn}},
  \bibinfo{author}{\bibfnamefont{J.~D.} \bibnamefont{Sau}}, \bibnamefont{and}
  \bibinfo{author}{\bibfnamefont{S.}~\bibnamefont{Das~Sarma}},
  \bibinfo{journal}{Phys. Rev. Lett.} \textbf{\bibinfo{volume}{105}},
  \bibinfo{pages}{077001} (\bibinfo{year}{2010}).

\bibitem[{\citenamefont{Oreg et~al.}(2010)\citenamefont{Oreg, Refael, and von
  Oppen}}]{bib:Oreg}
\bibinfo{author}{\bibfnamefont{Y.}~\bibnamefont{Oreg}},
  \bibinfo{author}{\bibfnamefont{G.}~\bibnamefont{Refael}}, \bibnamefont{and}
  \bibinfo{author}{\bibfnamefont{F.}~\bibnamefont{von Oppen}},
  \bibinfo{journal}{Phys. Rev. Lett.} \textbf{\bibinfo{volume}{105}},
  \bibinfo{pages}{177002} (\bibinfo{year}{2010}).

\bibitem[{\citenamefont{Alicea}(2010)}]{bib:AliceaPRB}
\bibinfo{author}{\bibfnamefont{J.}~\bibnamefont{Alicea}},
  \bibinfo{journal}{Phys. Rev. B} \textbf{\bibinfo{volume}{81}},
  \bibinfo{pages}{125318} (\bibinfo{year}{2010}).

\bibitem[{\citenamefont{Klinovaja
  et~al.}(2012{\natexlab{a}})\citenamefont{Klinovaja, Gangadharaiah, and
  Loss}}]{bib:MF_CNT_2012}
\bibinfo{author}{\bibfnamefont{J.}~\bibnamefont{Klinovaja}},
  \bibinfo{author}{\bibfnamefont{S.}~\bibnamefont{Gangadharaiah}},
  \bibnamefont{and} \bibinfo{author}{\bibfnamefont{D.}~\bibnamefont{Loss}},
  \bibinfo{journal}{Phys. Rev. Lett.} \textbf{\bibinfo{volume}{108}},
  \bibinfo{pages}{196804} (\bibinfo{year}{2012}{\natexlab{a}}).

\bibitem[{\citenamefont{Klinovaja
  et~al.}(2012{\natexlab{b}})\citenamefont{Klinovaja, Ferreira, and
  Loss}}]{bib:bilayer_MF_2012}
\bibinfo{author}{\bibfnamefont{J.}~\bibnamefont{Klinovaja}},
  \bibinfo{author}{\bibfnamefont{G.~J.} \bibnamefont{Ferreira}},
  \bibnamefont{and} \bibinfo{author}{\bibfnamefont{D.}~\bibnamefont{Loss}},
  \bibinfo{journal}{Phys. Rev. B} \textbf{\bibinfo{volume}{86}},
  \bibinfo{pages}{235416} (\bibinfo{year}{2012}{\natexlab{b}}).

\bibitem[{\citenamefont{Klinovaja and Loss}(2013)}]{bib:nanoribbon_KL}
\bibinfo{author}{\bibfnamefont{J.}~\bibnamefont{Klinovaja}} \bibnamefont{and}
  \bibinfo{author}{\bibfnamefont{D.}~\bibnamefont{Loss}},
  \bibinfo{journal}{Phys. Rev. X} \textbf{\bibinfo{volume}{3}},
  \bibinfo{pages}{011008} (\bibinfo{year}{2013}).

\bibitem[{\citenamefont{Goldstein and Chamon}(2011)}]{bib:Chamon}
\bibinfo{author}{\bibfnamefont{G.}~\bibnamefont{Goldstein}} \bibnamefont{and}
  \bibinfo{author}{\bibfnamefont{C.}~\bibnamefont{Chamon}},
  \bibinfo{journal}{Phys. Rev. B} \textbf{\bibinfo{volume}{84}},
  \bibinfo{pages}{205109} (\bibinfo{year}{2011}).

\bibitem[{\citenamefont{Budich et~al.}(2012)\citenamefont{Budich, Walter, and
  Trauzettel}}]{bib:Trauzettel}
\bibinfo{author}{\bibfnamefont{J.~C.} \bibnamefont{Budich}},
  \bibinfo{author}{\bibfnamefont{S.}~\bibnamefont{Walter}}, \bibnamefont{and}
  \bibinfo{author}{\bibfnamefont{B.}~\bibnamefont{Trauzettel}},
  \bibinfo{journal}{Phys. Rev. B} \textbf{\bibinfo{volume}{85}},
  \bibinfo{pages}{121405} (\bibinfo{year}{2012}).

\bibitem[{\citenamefont{Rainis and Loss}(2012)}]{bib:Diego3}
\bibinfo{author}{\bibfnamefont{D.}~\bibnamefont{Rainis}} \bibnamefont{and}
  \bibinfo{author}{\bibfnamefont{D.}~\bibnamefont{Loss}},
  \bibinfo{journal}{Phys. Rev. B} \textbf{\bibinfo{volume}{85}},
  \bibinfo{pages}{174533} (\bibinfo{year}{2012}).

\bibitem[{\citenamefont{Schmidt et~al.}(2012)\citenamefont{Schmidt, Rainis, and
  Loss}}]{bib:Diego2}
\bibinfo{author}{\bibfnamefont{M.~J.} \bibnamefont{Schmidt}},
  \bibinfo{author}{\bibfnamefont{D.}~\bibnamefont{Rainis}}, \bibnamefont{and}
  \bibinfo{author}{\bibfnamefont{D.}~\bibnamefont{Loss}},
  \bibinfo{journal}{Phys. Rev. B} \textbf{\bibinfo{volume}{86}},
  \bibinfo{pages}{085414} (\bibinfo{year}{2012}).

\bibitem[{\citenamefont{Hekking and Nazarov}(1994)}]{bib:Hekking}
\bibinfo{author}{\bibfnamefont{F.~W.~J.} \bibnamefont{Hekking}}
  \bibnamefont{and} \bibinfo{author}{\bibfnamefont{Y.~V.}
  \bibnamefont{Nazarov}}, \bibinfo{journal}{Phys. Rev. B}
  \textbf{\bibinfo{volume}{49}}, \bibinfo{pages}{6847} (\bibinfo{year}{1994}).

\bibitem[{\citenamefont{Deutscher and Feinberg}(2000)}]{bib:Deutscher}
\bibinfo{author}{\bibfnamefont{G.}~\bibnamefont{Deutscher}} \bibnamefont{and}
  \bibinfo{author}{\bibfnamefont{D.}~\bibnamefont{Feinberg}},
  \bibinfo{journal}{App. Phys. Lett.} \textbf{\bibinfo{volume}{76}},
  \bibinfo{pages}{487} (\bibinfo{year}{2000}).

\bibitem[{\citenamefont{Falci et~al.}(2001)\citenamefont{Falci, Feinberg, and
  Hekking}}]{bib:Falci}
\bibinfo{author}{\bibfnamefont{G.}~\bibnamefont{Falci}},
  \bibinfo{author}{\bibfnamefont{D.}~\bibnamefont{Feinberg}}, \bibnamefont{and}
  \bibinfo{author}{\bibfnamefont{F.~W.~J.} \bibnamefont{Hekking}},
  \bibinfo{journal}{EPL (Europhysics Letters)} \textbf{\bibinfo{volume}{54}},
  \bibinfo{pages}{255} (\bibinfo{year}{2001}).

\bibitem[{\citenamefont{Stanescu et~al.}(2011)\citenamefont{Stanescu, Lutchyn,
  and Das~Sarma}}]{DasSarmaPRB2011}
\bibinfo{author}{\bibfnamefont{T.~D.} \bibnamefont{Stanescu}},
  \bibinfo{author}{\bibfnamefont{R.~M.} \bibnamefont{Lutchyn}},
  \bibnamefont{and}
  \bibinfo{author}{\bibfnamefont{S.}~\bibnamefont{Das~Sarma}},
  \bibinfo{journal}{Phys. Rev. B} \textbf{\bibinfo{volume}{84}},
  \bibinfo{pages}{144522} (\bibinfo{year}{2011}).

\bibitem[{\citenamefont{Chevallier et~al.}(2013)\citenamefont{Chevallier,
  Simon, and Bena}}]{bib:PascalDisorder}
\bibinfo{author}{\bibfnamefont{D.}~\bibnamefont{Chevallier}},
  \bibinfo{author}{\bibfnamefont{P.}~\bibnamefont{Simon}}, \bibnamefont{and}
  \bibinfo{author}{\bibfnamefont{C.}~\bibnamefont{Bena}}
  (\bibinfo{year}{2013}), \bibinfo{note}{arXiv:1301.7420}.

\bibitem[{\citenamefont{Klinovaja and Loss}(2012)}]{bib:Lena}
\bibinfo{author}{\bibfnamefont{J.}~\bibnamefont{Klinovaja}} \bibnamefont{and}
  \bibinfo{author}{\bibfnamefont{D.}~\bibnamefont{Loss}},
  \bibinfo{journal}{Phys. Rev. B} \textbf{\bibinfo{volume}{86}},
  \bibinfo{pages}{085408} (\bibinfo{year}{2012}).

\bibitem[{\citenamefont{Zazunov and Egger}(2012)}]{bib:Zazunov}
\bibinfo{author}{\bibfnamefont{A.}~\bibnamefont{Zazunov}} \bibnamefont{and}
  \bibinfo{author}{\bibfnamefont{R.}~\bibnamefont{Egger}},
  \bibinfo{journal}{Phys. Rev. B} \textbf{\bibinfo{volume}{85}},
  \bibinfo{pages}{104514} (\bibinfo{year}{2012}).

\bibitem[{\citenamefont{Rainis et~al.}(2013)\citenamefont{Rainis, Trifunovic,
  Klinovaja, and Loss}}]{bib:Diego1}
\bibinfo{author}{\bibfnamefont{D.}~\bibnamefont{Rainis}},
  \bibinfo{author}{\bibfnamefont{L.}~\bibnamefont{Trifunovic}},
  \bibinfo{author}{\bibfnamefont{J.}~\bibnamefont{Klinovaja}},
  \bibnamefont{and} \bibinfo{author}{\bibfnamefont{D.}~\bibnamefont{Loss}},
  \bibinfo{journal}{Phys. Rev. B} \textbf{\bibinfo{volume}{87}},
  \bibinfo{pages}{024515} (\bibinfo{year}{2013}).

\bibitem[{\citenamefont{Alicea et~al.}(2011)\citenamefont{Alicea, Oreg, Refael,
  von Oppen, and Fisher}}]{bib:Alicea_Nphys}
\bibinfo{author}{\bibfnamefont{J.}~\bibnamefont{Alicea}},
  \bibinfo{author}{\bibfnamefont{Y.}~\bibnamefont{Oreg}},
  \bibinfo{author}{\bibfnamefont{G.}~\bibnamefont{Refael}},
  \bibinfo{author}{\bibfnamefont{F.}~\bibnamefont{von Oppen}},
  \bibnamefont{and} \bibinfo{author}{\bibfnamefont{M.~P.~A.}
  \bibnamefont{Fisher}}, \bibinfo{journal}{Nat. Phys.}
  \textbf{\bibinfo{volume}{7}}, \bibinfo{pages}{412} (\bibinfo{year}{2011}).

\end{thebibliography}

\end{document}